\documentclass[preprint,showpacs,superscriptaddress]{revtex4}

\usepackage{amsmath}
\usepackage{amssymb}
\usepackage{amscd}
\usepackage{wasysym}
\usepackage[ansinew]{inputenc}
\usepackage[T1]{fontenc}
\usepackage{ae,aecompl}

\usepackage[dvips]{graphicx}
\DeclareGraphicsExtensions{.eps}

\newcommand{\be}{\begin{equation}}
\newcommand{\ee}{\end{equation}}
\newcommand{\bea}{\begin{eqnarray}}
\newcommand{\eea}{\end{eqnarray}}
\newcommand{\nn}{\nonumber}

\renewcommand{\vec}[1]{\mathbf{#1}}

\newcommand{\tr}{{\rm tr}}

\begin{document}
\bibliographystyle{apsrev}

\title{Dissipation induced coherence of a two-mode Bose-Einstein condensate}
\author{D. Witthaut}
\email{dirk.witthaut@nbi.dk}
\affiliation{QUANTOP, Niels Bohr Institute, University of Copenhagen, 
DK--2100 Copenhagen, Denmark}
\author{F. Trimborn}
\affiliation{Institut f\"ur mathematische Physik, TU Braunschweig, D--38106 Braunschweig, Germany}
\author{S. Wimberger}
\affiliation{Institut f\"ur theoretische Physik, Universit\"{a}t Heidelberg, D--69120, Heidelberg, Germany}
\date{\today }

\begin{abstract}
We discuss the dynamics of a Bose-Einstein condensate in a 
double-well trap subject to phase noise and particle loss. 
The phase coherence of a weakly-interacting condensate as 
well as the response to an external driving show a pronounced 
stochastic resonance effect: Both quantities become maximal 
for a finite value of the dissipation rate matching the intrinsic 
time scales of the system. 
Even stronger effects are observed when dissipation acts in 
concurrence with strong inter-particle interactions, restoring
the purity of the condensate almost completely and increasing
the phase coherence significantly.
\end{abstract}

\maketitle


In our naive understanding thermal noise is generally deconstructive,
hindering measurements and degrading coherences in quantum mechanics.
A paradigmatic counterexample to this assertion is the effect of 
stochastic resonance (SR), where the response of a system to an external 
driving assumes its maximum in the presence of a finite amount of 
thermal noise \cite{Benz81}.
This maximum occurs when the time scales of the noise and 
the driving match. By now, SR has been shown in a variety of systems, 
an overview is given in the review articles 
\cite{Wies95,Dykm95,Gamm98,Well04}.

In addition to numerous examples in classical dynamics, SR has 
also been found in a variety of quantum systems (see \cite{Well04}
and references therein). Recently, 
there has been an increased interest in controlling and even exploiting 
dissipation in interacting many-body quantum systems. For instance, 
the entanglement in a spin chain assumes an SR-like maximum for a 
finite amount of thermal noise \cite{Huel07}.
Furthermore, it has been shown that dissipative processes can be 
tailored to prepare arbitrary pure states for quantum computation and 
strongly correlated states of ultracold atoms \cite{Krau08} or to 
implement a universal set of quantum gates \cite{Vers08}.
Actually, a recent experiment has even proven that strong inelastic 
collisions may inhibit particle losses and induce strong 
correlations in a quasi one-dimensional gas of ultracold atoms 
\cite{Syas08}.

In this letter we demonstrate the constructive effects of dissipation such 
as stochastic resonance for an interacting many-particle quantum system
realized by ultracold atoms in a double-well trap with biased particle
dissipation. It is shown that a proper amount of dissipation increases 
the coherence of the atomic cloud, especially in concurrence with strong 
inter-particle interactions.
These effects are of considerable strength for realistic parameters
and thus should be observable in ongoing experiments 
\cite{Albi05,Gati06,Schu05b,Foll07}.

The unitary dynamics of ultracold atoms in a double-well trap is described 
by the two-mode Bose-Hubbard Hamiltonian \cite{Milb97,Smer97,Angl01}
\bea
  \hat H &=&
     -J \left( \hat a_1^\dagger \hat a_2 +  \hat a_2^\dagger \hat a_1 \right)
     + \epsilon (\hat n_2 - \hat n_1)  \nn \\
     && \quad + \frac{U}{2} \left( \hat n_1(\hat n_1 -1) + \hat n_2(\hat n_2 -1)
    \right),
    \label{eqn-hami-bh}
\eea
where $\hat a_j$ and $\hat a_j^\dagger$ are the bosonic 
annihilation and creation operators in the $j$th well and 
$\hat n_j = \hat a_j^\dagger \hat a_j$ are the number 
operators. In general we consider a biased double-well trap, 
where the ground state energies of the two wells differ by 
$2\epsilon$. We set $\hbar = 1$, thus measuring all energies 
in frequency units.

The main source of decoherence is phase noise due to elastic collisions 
with atoms in the thermal cloud \cite{Angl97,Ruos98} which effectively 
heats the system. The heating rate is fixed as $\gamma_p = 5 \,{\rm s}^{-1}$
in the following, which is a realistic value for the experiments in Heidelberg 
\cite{Albi05,Gati06}. Methods to attenuate this source of decoherence were 
discussed only recently \cite{Khod08}.
Amplitude noise, i.e.~the exchange of particles with 
the thermal cloud due to inelastic scattering, drives the system to 
thermal equilibrium. However, this effect is usually much too weak to
produce the effects discussed below in present experiments (cf.~the 
discussion in \cite{Ruos98}).
In contrast, a strong and tunable source of dissipation can be implemented 
artificially by shining a resonant laser beam onto the trap, that removes
atoms with the site-dependent rates $\gamma_{aj}$ from the two wells $j = 1,2$. 
Non-trivial effects of dissipation such as the stochastic resonance discussed 
below require strongly biased loss rates, i.e. $\gamma_{a1} \neq \gamma_{a2}$. 
For a laser beam 
focused on one of the wells an asymmetry of
$f_a = (\gamma_{a2}$ -- $\gamma_{a1})/ (\gamma_{a2}$ + $\gamma_{a1}) = 0.5$
should be feasible.
Thus we consider the dynamics generated by the master equation
\bea
  \dot{\hat \rho} &=& -i [\hat H,\hat \rho] 
  - \frac{\gamma_p}{2} \sum_{j = 1,2}
    \left( \hat n_j^2 \hat \rho + \hat \rho \hat n_j^2 
          - 2 \hat n_j \hat \rho \hat n_j  \right) \nn \\
   && \quad - \frac{1}{2}  \sum_{j=1,2}  \gamma_{aj} \left(
     \hat a_j^\dagger \hat a_j \hat \rho + \hat \rho \hat a_j^\dagger \hat a_j  
    - 2 \hat a_j \hat \rho \hat a_j^\dagger \right).
   \label{eqn-master2}
\eea
The macroscopic dynamics of the atomic cloud is well described by a 
mean-field approximation, considering only the expectation values 
$s_j = 2 \, \tr(\hat L_j \hat \rho)$ of the angular momentum operators
\bea
  &&\hat L_x = \frac{1}{2} 
      \left( \hat a_1^\dagger \hat a_2  + \hat a_2^\dagger \hat a_1 \right), \quad 
  \hat L_y = \frac{i}{2} 
      \left( \hat a_1^\dagger \hat a_2  - \hat a_2^\dagger \hat a_1 \right),
      \nn \\
  && \qquad \qquad \quad \hat L_z = \frac{1}{2} 
      \left( \hat a_2^\dagger \hat a_2  - \hat a_1^\dagger \hat a_1 \right)
  \label{eqn-angular-op} 
\eea
and the particle number $n = \tr((\hat n_1 + \hat n_2) \hat \rho)$. The time
evolution of the Bloch vector $\vec s$ and the particle number is then
given by 
\bea
  \dot s_x &=& -2 \epsilon s_y - U s_y s_z - T_2^{-1} s_x, \nn \\
  \dot s_y &=&  2 J s_z + 2\epsilon  s_x + U s_x s_z  - T_2^{-1} s_y, \nn \\
  \dot s_z &=& - 2 J s_y - T_1^{-1} s_z - T_1^{-1} f_a n, \nn \\
  \dot n   &=& - T_1^{-1} n  - T_1^{-1} f_a s_z.
  \label{eqn-eom-bloch}
\eea
As usual expectation values of products have been factorized in the
$U$-dependent interaction terms to obtain a closed set of evolution equations 
\cite{Milb97,Smer97,Angl01}, whereas the dissipation terms are exact.
Furthermore we have defined the transversal and longitudinal damping times
by
\be
  T_1^{-1} = (\gamma_{a1} + \gamma_{a2})/2 \quad \mbox{and} \quad
  T_2^{-1} = \gamma_p +  T_1^{-1}.
  \label{eqn-rel-times}
\ee
These equations of motion resemble the celebrated Bloch equations in 
nuclear magnetic resonance \cite{Bloc46,Viol00} with some subtle but 
nevertheless important differences. The longitudinal relaxation is now
associated with particle loss and, more important, the dynamics is
substantially altered by the interaction term \cite{Milb97,Smer97,Albi05}.

\begin{figure}[tb]
\centering
\includegraphics[width=8cm, angle=0]{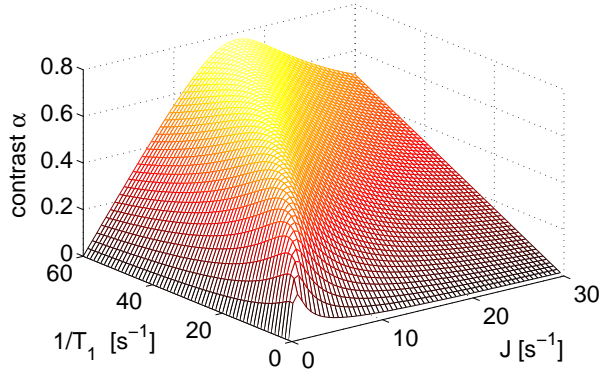}
\caption{\label{fig-contrast-3d}
(Color online) Contrast $\alpha$ in the quasi-steady state in 
dependence on the tunneling rate $J$ and the dissipation rate $1/T_1$ for
$U = 0$ and $\epsilon = 0$.
}
\end{figure}

In the following we will show that a finite amount of dissipation
induces a maximum of the coherence which can be understood as an 
SR effect.
In this discussion we have to distinguish between two different kinds
of coherence, which will both be considered in the following. First of 
all we consider the phase coherence between the two wells, which is 
measured by the average \textit{contrast} in interference experiments 
as described in \cite{Albi05,Gati06} and given by
\be
  \alpha = \frac{2|\langle \hat a_1^\dagger \hat a_2\rangle|
            }{\langle \hat n_1 + \hat n_2 \rangle } 
         = \frac{\sqrt{s_x^2+s_y^2}}{n} \, .
  \label{eqn-alpha-def}
\ee
Secondly, we will analyze how close the many-particle quantum state
is to a pure Bose-Einstein condensate (BEC), which is a coherent state for
the $SU(2)$ operator algebra \cite{08phase}. This property is quantified 
by the purity $p = 2 \, \tr(\hat \rho_{\rm red}^2)-1 = |\vec s|^2/n^2$ 
of the reduced single-particle density matrix $\hat \rho_{\rm red}$, 
cf.~\cite{Angl01}.

\begin{figure}[tb]
\centering
\includegraphics[width=7cm, angle=0]{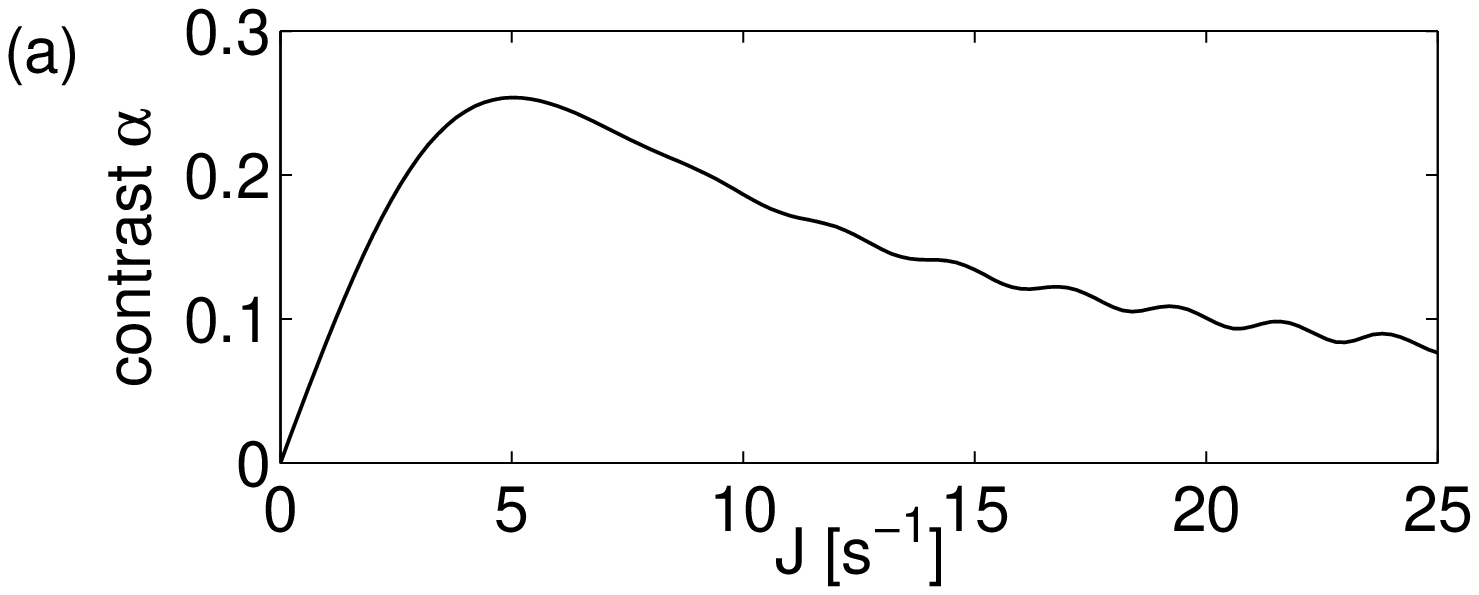}
\includegraphics[width=8cm, angle=0]{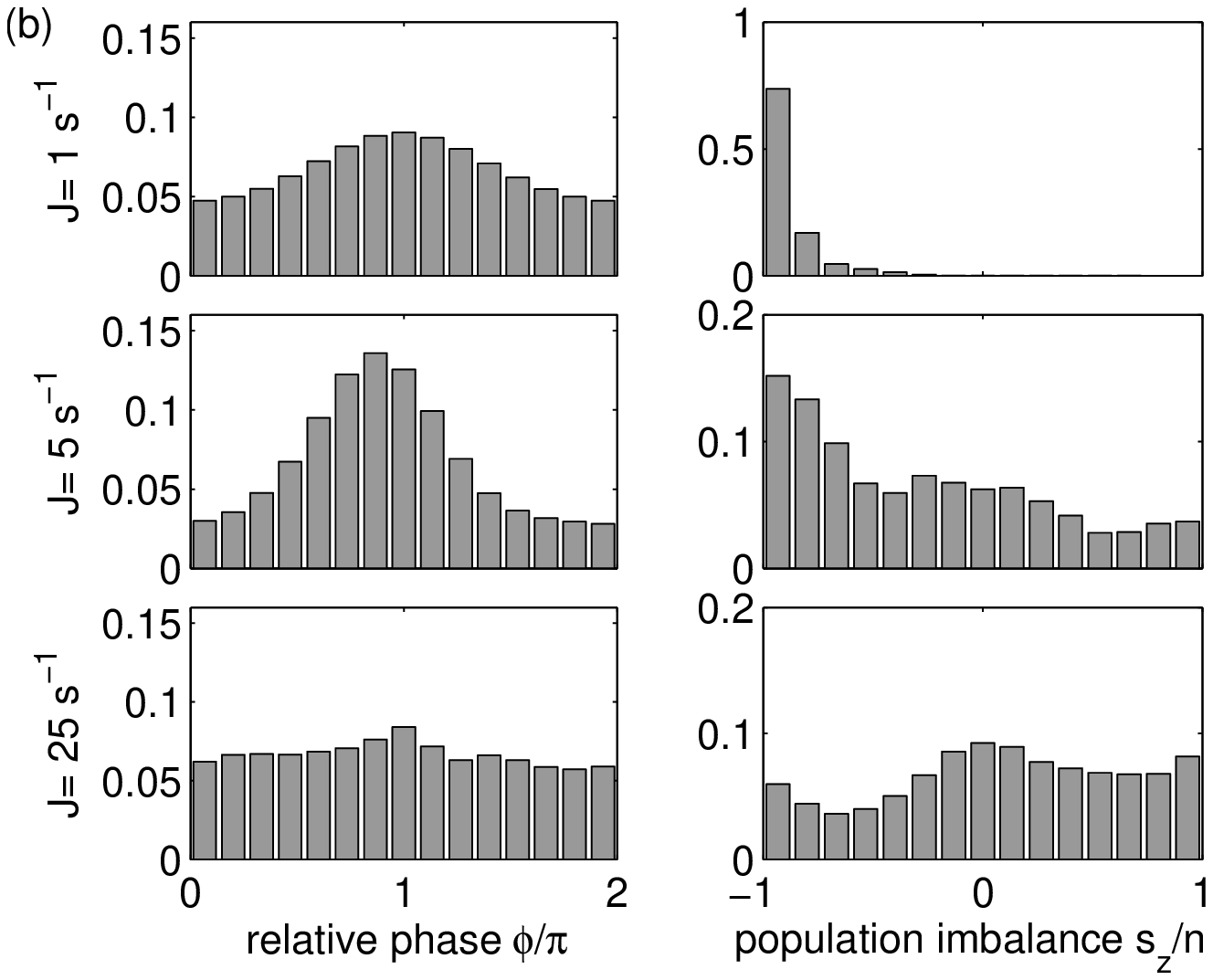}
\caption{\label{fig-qjump-lin}
(a) Average contrast $\alpha$ after $1.5 \, {\rm s}$ of propagation starting 
from a pure BEC (i.e.~a product state) with $s_z = n/2$ and $n(0) = 100$ 
particles in dependence of the tunneling rate $J$ for $T_1 = 0.5 \, {\rm s}$, 
$\epsilon = 10 \, {\rm s}^{-1}$, $U=0.1 \, {\rm s}^{-1}$.
(b) Histogram of the probabilities to measure the relative phase $\phi$
and the relative population imbalance $s_z/n$ in a single experimental 
run after $t = 1.5 \, {\rm s}$ obtained from a MCWF simulation of the 
many-body dynamics.
}
\end{figure}

Let us first discuss the weakly-interacting case, where the 
mean-field equations of motion (\ref{eqn-eom-bloch}) provide 
an excellent description of the dynamics, which is exact
for $U=0$. Obviously, only the trivial solution $\vec s = 0$ and 
$n = 0$ is a steady state in the strict sense.
However, the system rapidly relaxes to a quasi-steady state where the 
internal dynamics is completely frozen out and all components of the 
Bloch vector and the particle number decay at the same rate. 
Fig.~\ref{fig-contrast-3d} shows the contrast $\alpha$ for this quasi-steady 
state as a function of the tunneling rate $J$ and the dissipation rate $1/T_1$
for $U=0$. For a fixed value of one of the parameters, say $J$, one observes 
a typical SR-like maximum of the contrast for a finite value of the dissipation
rate $1/T_1$. In particular, the contrast is maximal if the time scales of the 
tunneling and the dissipation are matched according to \cite{08srlong}
\be
  f_a T_1^{-1} \approx 2J.
  \label{eqn-sr-condition}
\ee
This scenario is robust and not altered by weak inter-particle
interactions. Changes in the system parameters such as $\epsilon$ 
preserve the general shape of $\alpha(1/T_1,J)$ and the existence of a 
pronounced SR-like maximum. At the most, the function $\alpha(1/T_1,J)$ 
is stretched, shifting the position of the SR-like maximum.

The occurrence of a maximum of the contrast is explained by 
Fig.~\ref{fig-qjump-lin} (b), where the results of a Monte 
Carlo wave function (MCWF) simulation \cite{Dali92} of the 
many-body dynamics are shown for three different values of 
$J$ and $U=0.1 \, {\rm s}^{-1}$. We have plotted a histogram 
of the probabilities to observe the relative population 
imbalance $s_z$ and the relative phase $\phi$ in a single 
experimental run for three different values of the tunneling 
rate $J$ after the system has relaxed to the quasi-steady state.
With increasing $J$, the atoms are distributed more equally between
the two wells so that the single shot contrast increases.
Within the mean-field description this is reflected
by an increase of $\sqrt{s_x^2+s_y^2}/|\vec s|$ at the 
expense of $|s_z|$.
However, this effect also makes 
the system more vulnerable to phase noise so that the relative phase 
of the two modes becomes more and more random and $|\vec s|/n$ 
decreases. The average contrast (\ref{eqn-alpha-def}) then assumes a 
maximum for intermediate values of $J$ as shown in 
Fig.~\ref{fig-qjump-lin} (a). 
Note that the trap is assumed to be weakly biased in this example, 
shifting the position of the SR-like maximum to a value of $J$ which 
is easier accessible in ongoing experiments \cite{Albi05,Gati06}.

\begin{figure}[tb]
\centering
\includegraphics[width=8cm, angle=0]{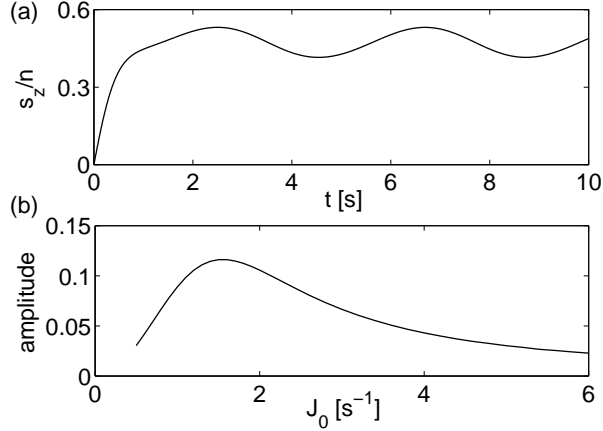}
\caption{\label{fig-sr-driven}
(a) Oscillation of the relative population imbalance $s_z/n$ of
a weakly driven two-mode BEC for $J_0 = 1.5 \, {\rm s}^{-1}$, 
$T_1 = 0.5 \, {\rm s}$ and $\epsilon = 0$.
(b) Amplitude of the oscillations in dependence on 
the tunneling rate $J_0$.
}
\end{figure}

So far we have demonstrated a stochastic resonance of the contrast 
for a BEC in a static double-well trap with biased particle losses.
In the following we will show that the system's response to a weak 
external driving also assumes a maximum for a finite dissipation 
rate -- an effect which is conceptually closer to the common 
interpretation of stochastic resonance. 
We consider a weak driving of the tunneling rate
$J(t) = J_0 + J_1 \cos(\omega t)$ at the resonance frequency 
$\omega = \sqrt{J_0^2 + \epsilon^2}$, where the amplitude is not
more than $J_1/J_0 = 10\%$. This can be readily implemented in 
optical setups by varying the intensity of the counter-propagating 
lasers forming the optical lattice. 
Fig.~\ref{fig-sr-driven} (a) shows the resulting dynamics for 
$T_1 = 0.5 \, {\rm s}$ and $J_0 = 1.5 \, {\rm s}^{-1}$. After a short 
transient period, the relative population imbalance $s_z(t)/n(t)$ 
oscillates approximately sinusoidally. The system response measured 
by the amplitude of these forced oscillations shows the familiar 
SR-like maximum as illustrated in Fig.~\ref{fig-sr-driven} (b).
It should be detectable without major problems in ongoing 
experiments, in which the population imbalance $s_z$ can be 
measured with a resolution of a few atoms \cite{Albi05,Gati06}. 
A more detailed study of the parameter ranges for such a driven
case of SR will be discussed in a forthcoming article 
\cite{08srlong}.

\begin{figure}[tb]
\centering
\includegraphics[width=8cm, angle=0]{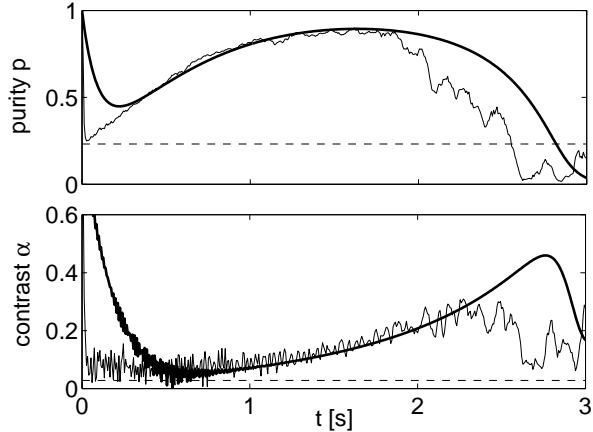}
\caption{\label{fig-dyn-qjump}
Time evolution of the purity $p$ and the contrast $\alpha$ for
$J=U=10 \, {\rm s}^{-1}$, $\epsilon= 0$, $T_1 = 0.5 \, {\rm s}$.
The initial state is a pure BEC  with $s_z = n/2$ and 
$n(0) = 100$ particles. 
The results of a MCWF simulation averaged over 100 runs
are plotted as a thin solid line while the mean-field results are 
plotted as a thick line. The dashed line shows the steady state values 
for $1/T_1 = 1/T_2 = 0$, i.e. without coupling to the environment.
}
\end{figure}

Even more remarkable values of the coherences are observed in the
case of strong interactions, which is experimentally most relevant 
and theoretically most profound. The interplay between interactions 
and dissipation significantly increases the coherences in comparison 
to situations where one of the two is weak or missing.
An example for the dynamics of a strongly-interacting BEC is shown in Fig.~\ref{fig-dyn-qjump} for an initially pure BEC with $s_z = n/2$,
calculated both with the MCWF method and within the mean-field 
approximation (\ref{eqn-eom-bloch}). 
At first, the purity $p$ and the contrast $\alpha$ drop rapidly 
due to the phase noise and the interactions, cf.~\cite{Angl01}.
For intermediate times, however, the system relaxes to a nonlinear 
quasi-steady state, which is a nearly pure BEC mostly localized in 
the well with the smaller decay rate. Consequently, the purity $p$ 
is restored almost completely and the contrast $\alpha$ is relatively 
large. 
In close analogy to the celebrated self-trapping effect 
\cite{Milb97,Smer97,Albi05}, this quasi-steady state exists only 
as long as the effective interaction strength $Un(t)$ is larger than 
a critical value given by \cite{08mfdecay}
\be
  U^2 n^2 \apprge 4J^2 - f_a^2 T_1^{-2} \, .
\label{eqn-critU}
\ee
As the particle number $n$ decays, this state ceases to exist so that 
the system relaxes to a linear quasi-steady state with considerably
smaller values of $p$ and $\alpha$ as discussed above for the 
weakly-interacting case. 

\begin{figure}[tb]
\centering
\includegraphics[width=8cm, angle=0]{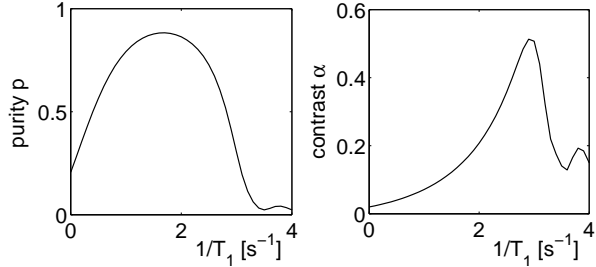}
\caption{\label{fig-coh-5s}
Purity $p$ and contrast $\alpha$ after $t = 2 \, {\rm s}$
in dependence on the dissipation rate $1/T_1$ calculated 
within the mean-field approximation. The remaining parameters 
are chosen as in Fig.~\ref{fig-dyn-qjump}.
}
\end{figure}

Moreover, the coherences at intermediate times are also larger than 
in an interacting, but non-dissipative system. The dashed lines in  Fig.~\ref{fig-dyn-qjump} show the steady state values of the 
purity $p$ and the contrast $\alpha$ for $1/T_1 = 1/T_2 = 0$,
apart from occasional revivals due to the finite particle number.
It is observed that the coherences are considerably smaller
compared to the strongly-interacting open system. This loss of 
coherence can be understood by the fact that the interactions lead 
to an effective decoherence on the single-particle level 
\cite{Angl01}, degrading $\alpha$ and $p$. This effect is
mostly cured by the dissipation.

The behaviour illustrated in Fig.~\ref{fig-dyn-qjump} and 
discussed above is universal in the sense that the maxima of the 
purity and the contrast are present for all values of $U$ and
$1/T_1$ if only $Un(t=0)$ is well above the critical value 
(\ref{eqn-critU}) for the existence of the nonlinear quasi-steady 
state. However, the maxima occur later if $T_1$ or $U$ increase.
The purity $p$ and the contrast $\alpha$ after a fixed time 
$t = 2 \, {\rm s}$ are plotted in Fig.~\ref{fig-coh-5s} in 
dependence on the dissipation rate $1/T_1$, 
showing pronounced maxima for finite values of $1/T_1$.
For smaller dissipation rates, the maximum of the contrast has not 
been assumed yet while the system has already relaxed to the linear
quasi-steady state for larger values of $1/T_1$.

To summarize, we have shown that the coherence properties of a weakly 
and in particular also of a strongly interacting Bose-Einstein 
condensate in a double-well trap can be controlled by engineering 
the system's parameters and dissipation simultaneously. 
An important conclusion is that the interplay of interactions and 
dissipation can drive the system to a state of maximum coherence, 
while both processes alone usually lead to a loss of coherence.
Since the double-well BEC is nowadays routinely realized with nearly 
perfect control on atom-atom interactions and external potentials 
\cite{Albi05,Gati06}, we hope for an experimental verification and 
future extensions of the predicted stochastic resonance scheme.

We thank M.~K.~Oberthaler, J.~R.~Anglin and A.~S.~S\o{}rensen
for stimulating discussions.
This work has been supported by the German Research Foundation (DFG) 
through the research fellowship program (grant number WI 3415/1) and 
the Heidelberg Graduate School of Fundamental Physics (grant number 
GSC 129/1) as well as the Studienstiftung des deutschen Volkes.


\end{document}